\documentclass[prd,aps,floats,twocolumn,showpacs]{revtex4}
\usepackage[dvips]{epsfig}

\begin{document}

\def\sqr#1#2{{\vcenter{\hrule height.#2pt
   \hbox{\vrule width.#2pt height#1pt \kern#1pt
      \vrule width.#2pt}
   \hrule height.#2pt}}}
\def\square{{\mathchoice\sqr64\sqr64\sqr{3.0}3\sqr{3.0}3}}

\title{Local gauge invariance of free fields}

\author{Bernd A. Berg}

\affiliation{Department of Physics, Florida State
             University, Tallahassee, FL 32306-4350, USA}

\date{August 6, 2012.} 

\begin{abstract}
It is noted that, in contrast to widespread believes, free fields
do not only allow for global, but also for local gauge invariance.
\end{abstract}
\maketitle

Euclidean notation is used in this paper.
Let us consider the Lagrangian of a free fermion field
\begin{equation} \label{L_F}
  L_F\ =\ \overline{\psi}\left(\gamma_{\mu}\partial_{\mu}+m\right)\psi\,,
\end{equation}
which is invariant under the global symmetry transformations
\begin{eqnarray} \label{psipg}
  \psi\to\psi' = \exp[i\alpha]\,\psi\,,~~
  \overline{\psi}\to\overline{\psi}' = 
  \exp[-i\alpha]\,\overline{\psi}\,.
\end{eqnarray}
The interaction with vector gauge fields is often claimed to be 
a consequence of promoting (\ref{psipg}) to a local symmetry 
transformation
\begin{eqnarray} \label{psip}
  \psi\to\psi' = \exp[i\alpha(x)]\,\psi\,,~~
  \overline{\psi}\to\overline{\psi}' = 
  \exp[-i\alpha(x)]\,\overline{\psi}\,.
\end{eqnarray}
The QED Lagrangian
\begin{equation} \label{L_QED}
  L_{QED}\ =\ \frac{1}{4}\,F_{\mu\nu}F_{\mu\nu}
       + \overline{\psi}\left(\gamma_{\mu}D_{\mu}+m\right)\psi\,,
\end{equation}
with $F_{\mu\nu}=\partial_{\mu}A_{\nu}-\partial_{\nu}A_{\mu}$,
$D_{\mu}=\partial_{\mu}+igA_{\mu}$, and
\begin{equation} \label{Ap}
  A_{\mu}\to A'_{\mu} = A_{\mu}-g^{-1}\,\partial_{\mu}\alpha(x)\,,
\end{equation}
is invariant under the transformations (\ref{psip}) to (\ref{Ap}), 
but whether there is an interaction or not depends also on the measure 
used to define the QFT. Let us use instead of the QED partition function
\begin{eqnarray} \label{Z_QED}
  Z_{QED} = \int DA\,D\overline{\psi}\,D\psi\,\exp\left[ 
            \int d^4x\,L_{QED}\right]
\end{eqnarray}
the partition function
\begin{eqnarray} \label{Z_F}
  Z_F = \int D\alpha\,D\overline{\psi}\,D\psi\,\exp\left[\int d^4x\,
  \overline{\psi}\,(\gamma_{\mu}D^0_{\mu}+m)\,\psi\right]
\end{eqnarray}
where  the functional integration is over a scalar field $\alpha$ and
\vskip -20pt
\begin{eqnarray} \label{D0}
  D^0_{\mu}=\partial_{\mu}+i \partial_{\mu}\alpha(x)\,.
\end{eqnarray}
There is still no interaction, while local gauge invariance (\ref{psip}) 
is preserved. With $A_{\mu}=-g^{-1}\partial_{\mu}\alpha(x)$ we could 
as well use $L_{QED}$ in (\ref{Z_F}), because $F_{\mu\nu}F_{\mu\nu}=0$ 
holds for a pure gauge.

Statements from textbooks like \cite{peskin} ``Even the very existence 
of the vector field $A_{\mu}$ is a consequence of local symmetry: 
Without it we could not write an invariant Lagrangian involving 
derivatives of $\psi$." ought to be modified. If one wants to introduce 
an interaction and keep the local invariance, it has to be a gauge 
field. The local invariance does not enforce an interaction.

This observation might be of some physical relevance. Using for U(1) 
and SU(2) gauge fields a more elaborated construction that exploits 
integration over pure gauges allows one to extend gauge invariance, so 
that new invariant field combinations emerge, which generate in the
lattice regularization a mass for the SU(2) vector bosons \cite{berg12},
while direct SU(2) mass terms are still forbidden, because they break 
the extended gauge invariance as well as the usual. Whether a quantum 
continuum limit exists remains under investigation. One motivation for 
the present note is to give a simple example which illuminates extended 
gauge invariance used in~\cite{berg12}.

The above construction holds also for complex scalar bosons for which
one can easily implement lattice field theory simulations. With the 
local gauge transformations
\begin{eqnarray} \label{phip}
  \phi\to\phi' = \exp[i\alpha(x)]\,\phi\,,~~
  \overline{\phi}\to\overline{\phi}' = 
  \exp[-i\alpha(x)]\,\overline{\phi}
\end{eqnarray}
Lagrangian and partition function are
\begin{eqnarray} \label{L_B}
  L_B &=& \left|D^0_{\mu}\phi(x)\right|^2 + 
          m^2\,\overline{\phi}(x)\,\phi(x) \\ \label{Z_B}
  Z_B &=& \int D\alpha\,D\overline{\phi}\,D\phi\,\exp\left[\int d^4x\,
  L_B\,\psi\right]\,.
\end{eqnarray}

In summary, the partition functions $Z_F$ (\ref{Z_F}) and $Z_B$ 
(\ref{Z_B}) describe free fermion and boson fields with local gauge 
invariance. 

\acknowledgments
This work was in part supported by the US Department of Energy under 
contract DE-FA02-97ER41022. This paper grew out of a report by an 
anonymous referee of Ref.~\cite{berg12}.

\clearpage
\end{document}